\documentclass[12pt,letterpaper]{article}
\pdfoutput=1
\usepackage[top=3cm, bottom=3cm, left=1cm, right=2cm]{geometry}
\usepackage[usenames,dvipsnames,svgnames]{xcolor}
\definecolor{darkblue}{rgb}{0.0,0.1,0.3} % dark blue
\definecolor{darkgreen}{rgb}{0,0.65,0}
\definecolor{dblue4}{rgb}{0.06,0.31,0.55} % DodgerBlue4
\definecolor{nicered}{rgb}{0.7,0.1,0.1}
\definecolor{nicegreen}{rgb}{0.1,0.5,0.1}

\usepackage[numbers,sort&compress]{natbib}
\usepackage[utf8x]{inputenc}
\usepackage{amsmath,amssymb,amsfonts,amsthm}
\usepackage{xcolor}
\usepackage[colorlinks=true,citecolor= nicegreen,linkcolor=nicered]{hyperref}

\usepackage{verbatim}
\usepackage{graphicx}
\graphicspath{{figures/}}
\usepackage{cancel}
\usepackage{tipa}
\usepackage{array}   % for \newcolumntype macro
\newcolumntype{L}{>{$}l<{$}} % math-mode version of "l" column type
\newcolumntype{R}{>{$}r<{$}} % math-mode version of "l" column type
\usepackage{booktabs}
\usepackage{tabularx}
\usepackage{multirow}
\usepackage{dsfont}
\newcolumntype{Y}{>{\centering\arraybackslash}X}
\usepackage{tikz}
\newcommand{\ReportNumbers}[1]{%
\begin{tikzpicture}[overlay, remember picture]
\path (current page.north east) ++(-1,-1) node[below left] {#1};
\end{tikzpicture}
}

\def\c{,\allowbreak}

\title{Anomaly-free Abelian gauge symmetries\\with Dirac seesaws}

\author{Nicolás Bernal\footnote{\href{mailto:nicolas.bernal@uan.edu.co}{nicolas.bernal@uan.edu.co}}\\
\textit{\small  Centro de Investigaciones, Universidad Antonio Nariño}\\
\textit{\small  Carrera 3 Este \# 47A-15, Bogotá, Colombia}\\
[4mm]
Diego Restrepo\footnote{\href{mailto:restrepo@udea.edu.co}{restrepo@udea.edu.co}}\\
\textit{\small Instituto de Física, Universidad de Antioquia}\\
\textit{\small  Calle 70 \# 52-21, Apartado Aéreo 1226, Medellín, Colombia}
}
\date{}
\begin{document}
\maketitle
\ReportNumbers{\footnotesize PI/UAN-2021-696FT}

\begin{abstract}
We perform a {\it systematic} analysis of Standard Model extensions with an additional anomaly-free gauge $U(1)$ symmetry, to generate tree-level Dirac neutrino masses.
An anomaly-free symmetry demands nontrivial conditions on the charges of the unavoidable new states.
An intensive scan was performed, looking for solutions generating neutrino masses by the type-I and type-II tree-level Dirac seesaw mechanism, via operators with dimension 5 and 6, that correspond to active or dark symmetries.
Special attention was paid to the cases featuring no extra massless chiral fermions or multicomponent dark matter with unconditional stability.
\end{abstract}

%%%%%%%%%%%%%%%%%%%%%%%%%%%%%%%%%%%%%%%%%%%%%%%%%%%%%%%%%%%%
%%%%%%%%%%%%%%%%%%%%%%%%%%%%%%%%%%%%%%%%%%%%%%%%%%%%%%%%%%%%
%%%%%%%%%%%%%%%%%%%%%%%%%%%%%%%%%%%%%%%%%%%%%%%%%%%%%%%%%%%%
\section{Introduction}

%Dirac
Despite its enormous success, the standard model (SM) has to be extended in order to account for neutrino masses and dark matter (DM).
Regarding neutrinos, the oscillation data is compatible with both Majorana or Dirac neutrino masses~\cite{Zyla:2020zbs}, with no strong preference for either of the two possibilities.
Even if most of the literature assumes that neutrinos are Majorana in nature (see, e.g., Ref.~\cite{Cai:2017jrq} for a review), the mass generation mechanism for Dirac neutrinos has recently received increased attention.

To explain Dirac neutrino masses, right-handed neutrinos (RHNs) have to be introduced.
Additionally, an extra local symmetry is also required to guarantee proper total lepton number conservation~\cite{Ma:2014qra}.
Even so, the required Yukawa couplings are typically very suppressed, of the order $\mathcal{O}\left(10^{-10}\right)$, if Dirac neutrino masses are induced directly from the SM Higgs mechanism~\cite{Calle:2018ovc, CentellesChulia:2019gic}.  Nevertheless, if the symmetry forbids the tree-level contribution driven by the SM Higgs, a Dirac-seesaw mechanism can be implemented~\cite{Gu:2016hxh, Yao:2018ekp}.
For example, the type-I Dirac-seesaw~\cite{Roncadelli:1983ty,Roy:1983be,CentellesChulia:2016rms} could appear in the context of anomaly-free gauge $U(1)_{B-L}$~\cite{Ma:2014qra} or $U(1)_R$~\cite{Jana:2019mez} symmetries.
Specific models  have been studied for producing neutrino masses via the type-II Dirac Seesaw~\cite{Gu:2006dc,Bonilla:2016zef} with both anomaly-free gauge $U(1)_{B-L}$ symmetries~\cite{Gu:2019yvw, Nanda:2019nqy} or with anomaly-free Abelian dark symmetries~\cite{Ma:2021szi}.
The extra chiral singlet fermions required to cancel out the anomalies can be part of a hidden sector with DM candidates, as in the case type-I~\cite{Bernal:2018aon, Gu:2019ogb} or type-II~\cite{Patra:2016ofq} Majorana seesaw.
At loop level, the heavy particles in the radiative seesaw can be fully associated to an Abelian gauge dark symmetry $U(1)_D$ with the lightest of them as DM candidate~\cite{Gu:2007ug, Batell:2010bp, Farzan:2012sa, Calle:2019mxn}, as well to an active Abelian gauge symmetry $U(1)_X$, like an $U(1)_{B-L}$~\cite{Calle:2018ovc, Bonilla:2018ynb, Calle:2019mxn, Abada:2021yot}.
The studies of one-loop Dirac neutrino masses have typically focused on finding specific anomaly-free solutions of this two kinds of symmetries, see, e.g., Refs.~\cite{Saad:2019bqf, Bonilla:2019hfb, Jana:2019mez, Jana:2019mgj, Escribano:2020iqq}.
Only few studies have performed systematic analysis of SM extensions with an additional anomaly-free gauge $U(1)$ symmetry to generate Dirac neutrino masses~\cite{Wong:2020obo, Bernal:2021ezl}.

Here we continue the effort of performing a systematic analysis of SM extensions, presenting a complete set of relevant anomaly-free solutions to the general problem of the generation of neutrino masses by the type-I and type-II tree-level Dirac seesaw mechanism, via dimension 5 and 6 operators.
A full set of relevant solutions is obtained.
Our method can be easily applied to find the full set of anomaly-free solutions to well defined phenomenological problems. 

In this work, we look for anomaly-free solutions to SM extensions with an additional $U(1)$ gauge symmetry, giving rise to tree-level Dirac neutrino masses.
For that purpose, in section~\ref{sec:anomaly} we briefly revise the conditions to have a non-anomalous $U(1)$ gauge symmetry.
In section~\ref{sec:Dirac} we discuss solutions giving rise to neutrino masses, by the type-I and type-II tree-level Dirac seesaw mechanism, via operators with dimensions 5 and 6, corresponding to active or dark symmetries.
Finally, in section~\ref{sec:con} our conclusions are presented.

%%%%%%%%%%%%%%%%%%%%%%%%%%%%%%%%%%%%%%%%%%%%%%%%%%%%%%%%%%%%
%%%%%%%%%%%%%%%%%%%%%%%%%%%%%%%%%%%%%%%%%%%%%%%%%%%%%%%%%%%%
%%%%%%%%%%%%%%%%%%%%%%%%%%%%%%%%%%%%%%%%%%%%%%%%%%%%%%%%%%%%
\section{Anomaly conditions} \label{sec:anomaly}
We consider an extension of the SM with an additional $U(1)_X$ gauge symmetry, and $N'$ right-handed chiral fields $\psi_\rho$ singlets under the SM $SU(3)_c\otimes SU(2)_L\otimes U(1)_Y$ group, with charges $n_{\rho}$ under the $U(1)_X$, where $\rho=1,\cdots,N'$.
Additionally, we assume that the SM right-handed chiral fermions transform under the {\it active} $U(1)_X$ symmetry, with charges denoted with the same name of the field.%
\footnote{$Q$ and $L$ are the $X$-charges of the fermion doublets $Q^{\dagger}$ and $L^{\dagger}$, respectively.
Also, for the hypercharges we have $Y_{L^\dagger} = +1$ and $Y_{Q^\dagger} = -\frac13$.}
To avoid having an anomalous $U(1)_X$, the three linear anomaly conditions to be fulfilled are
\begin{align}
    \left[SU(3)_c\right]^{2}  U(1)_{X} &: \quad [3u+3d] + [3\times 2Q]=0\,,\\
    \left[SU(2)_{L}\right]^{2} U(1)_{X} &: \quad [2L+3\times 2Q]=0\,,\\
    \left[U(1)_{Y}\right]^{2} U(1)_{X} &: \quad \left[(-2)^2e + 3\left(\frac43\right)^2u + 3\left(-\frac23\right)^2d\right] + \left[2(1)^2L + 3\times 2\left(-\frac13\right)^2Q\right] =0\,.
\end{align}
As they only depend on the SM fermions, three of their $X$-charges can be expressed in terms of the other two~\cite{Appelquist:2002mw, Campos:2017dgc, Das:2017flq, Calle:2019mxn}, chosen to be $e$ and $L$, as
\begin{align}\label{eq:sol0}
  u=&-e-\frac23 L\,,& d=& e+\frac43 L\,,& Q=& -\frac13 L\,.
\end{align}
We note that the quadratic anomaly condition in $U(1)_X$ is trivially satisfied. 
However, the mixed gauge-gravitational $\left[ \text{Grav} \right]^{2} U(1)_{X}$ and the cubic $\left[U(1)_{X}\right]^{3}$ anomalies do depend on the extra fermion charges $n_\rho$, and therefore two additional conditions have to be imposed in order to avoid an anomalous $U(1)_X$~\cite{Calle:2019mxn}:
\begin{align}\label{eq:grav}
  \sum_{\rho=1}^{N'}n_{\rho}+3m&=0\,,&  \sum_{\rho=1}^{N'}n_{\rho}^3+3m^3&=0\,,
\end{align}
where $m \equiv e+2L$.
Equation~\eqref{eq:sol0} can be conveniently rewritten as
\begin{align}
  u=&\frac{4 L}{3}-m\,,& d=m-\frac{2 L}{3}& \,,& Q=& -\frac{L}{3}\,,& e=& m-2 L\,.
\end{align}
Finally, we note that the SM Higgs must have an $X$-charge
\begin{align}
  h = -e-L = L-m
\end{align}
to guarantee that SM quarks and charged leptons acquire masses through the standard Higgs mechanism.%
\footnote{In particular, if $h=0$ a gauge symmetry with SM-fermion charge $X=m\,(B-L)$ is obtained, where $B-L$ are the baryon-minus-lepton charges.}
Along these lines, we also assume that the singlet chiral fermions $\psi_\rho$ only acquire mass through the spontaneous symmetry breaking (SSB) of the extra $U(1)_X$ symmetry.
This excludes solutions with vector-like states.
We note that the existence of fields charged under both hypercharge and $U(1)_X$ induce at loop level the kinetic mixing operator $\mathcal{L} \supset \frac{\epsilon}{2} B^{\mu\nu} X_{\mu\nu}$, where $B^{\mu\nu}$ and $X^{\mu\nu}$ are the field strengths related to the $U(1)_Y$ and the extra $U(1)_X$, respectively.
The dimensionless parameter $\epsilon$ depends on the masses of the particles in the loop, as well as their specific charge assignment and the gauge couplings under the two $U(1)$ symmetries~\cite{Holdom:1985ag, Cheung:2009qd, Gherghetta:2019coi}.

It is interesting to note that the conditions in Eq.~\eqref{eq:grav} are completely equivalent to the ones coming from a scenario where the SM is extended with a dark $U(1)_D$ gauge symmetry with $N=N'+3$ right-handed singlet chiral fermions, $N'$ of them with the charges $n_\rho$ and three with charge $m$, and where the SM is invariant (hence a {\it dark} symmetry)~\cite{Bernal:2021ezl}.
Even if comparable, there is a major technical advantage of the latter approach:
If the SM is extended with and additional dark $U(1)_D$ gauge symmetry (under which it is uncharged), and $N$ right-handed chiral fields singlets under the SM group, the $U(1)_D$ is not anomalous if the Diophantine equations
\begin{align} \label{eq:NN3}
 \sum_{\rho=1}^{N}n_{\rho}=0 \qquad \text{and} \qquad \sum_{\rho=1}^{N}n_{\rho}^3=0\,,
\end{align}
coming from the mixed gauge-gravitational $\left[ \text{Grav} \right]^{2} U(1)_D$ and cubic $\left[U(1)_D\right]^{3}$ conditions are fulfilled.

%%%%%%%%%%%%%%%%%%%%%%%%%%%%%%%%%%%%%%%%%%%%%%%%%%%%%%%%%%%%
%%%%%%%%%%%%%%%%%%%%%%%%%%%%%%%%%%%%%%%%%%%%%%%%%%%%%%%%%%%%
%%%%%%%%%%%%%%%%%%%%%%%%%%%%%%%%%%%%%%%%%%%%%%%%%%%%%%%%%%%%
\section{Dirac seesaw models} \label{sec:Dirac}
In this section we look for anomaly-free $U(1)_X$ gauge extensions of the SM, with  $N$ singlet chiral fermions, realizing the effective Dirac neutrino mass operators~\cite{Cleaver:1997nj, Gu:2006dc} at tree-level.
In the two-component spinor notation, they can be written as
\begin{align} \label{eq:nmo56}
    \mathcal{L}_{\text{eff}} = h_{\nu}^{\alpha i} \, \left( \nu_{R\alpha}\right)^{\dagger} \, \epsilon_{ab} \, L_i^a \, H^b \left(\frac{S^*}{\Lambda}\right)^\delta + \text{H.c.},\qquad \text{with $i=1,2,3$}\,,
\end{align}
and $\delta = 1$ or $2$ for dimension 5 (D-5) or 6 (D-6) operators, respectively.
Here $h_{\nu}^{\alpha i}$ correspond to dimensionless induced couplings, $\nu_{R\alpha}$ are at least two RHNs ($\alpha=1,2,\ldots$) with the same $X$-charge $\nu$, $L_{i}$ are the lepton doublets with $X$-charge $-L$, $H$ is the SM Higgs doublet with $X$-charge $h=L-m$, $S$ is the complex singlet scalar responsible for the SSB of the anomaly-free gauge symmetry with $X$-charge $s=-(\nu+m)/\delta$, respectively, and $\Lambda$ is a scale of new physics, which is parametrically the typical mass scale of the new (heavy) states.
In general, after the SSB, a remnant $\mathbb{Z}_{|s|}$ discrete symmetry is left, which guarantees the stability of a potential DM candidate~\cite{Batell:2010bp}.

%%%%%%%%%%%%%%%%%%%%%%%%%%%%%%%%%%%%%%%%%%%%%%%%%%
\subsection{Type-I Dirac Seesaw}

%%%%%%%%%%%%%%%%%%%%%%
\begin{figure}
  \def\scl{0.8}
  \centering
  \includegraphics[scale=\scl]{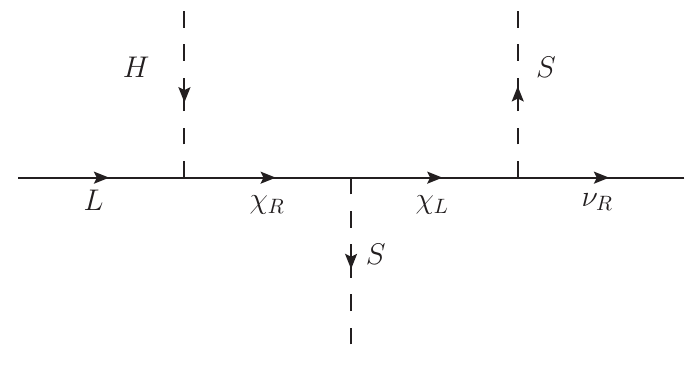}
  \caption{Diagram leading to tree-level type-I Dirac neutrino masses, via the dimension 6 operator described in Eq.~\eqref{eq:nmo56}.
 }
  \label{fig:topoI}
\end{figure}
%%%%%%%%%%%%%%%%%%%%%%
The realizations of type-I Dirac seesaw with D-5 operators automatically imply a vector-like pair of singlet fermions, and is therefore not possible with only chiral fermions.
Hence, we will not consider that case any longer in this study.
However, the situation is different for D-6 operators.
Figure~\ref{fig:topoI} presents the diagram realizing the D-6 effective Lagrangian in Eq.~\eqref{eq:nmo56} for type-I seesaw, if one only allows SSB masses for the singlet chiral fermions.
It is important to note that this process can only take place for {\it active} symmetries (a {\it dark} symmetry would imply a state $\chi_R$ with charge $r = 0$).
In that figure, the flux of charges in each vertex imply that
\begin{align} \label{eq:vertices}
    -m =& r\,, & r = & s-l\,,& -l = & s+\nu \,.
\end{align}
The charge of $S$ can be expressed as a function of $l$ and $m$ as
\begin{align}
    s = l-m\,,
\end{align}
and the chiral fermion charges obey
\begin{align} \label{eq:condition}
    \nu + 2l = m\,.
\end{align}
It is important to note that the first condition in Eq.~\eqref{eq:vertices} does {\it not} imply vector-like fermions, even if it contains particles with opposite charges, since the charge $m$ is associated to SM doublets. 
In this way, for any chiral solution (i.e., without opposite charges) we have two minimal ways of building a solution satisfying the condition in Eq.~\eqref{eq:condition}:
\begin{itemize}
    \item If $m$ is already present in the solution as a non-repeated $X$-charge $(m\c\ldots)$, we can add two sets of opposite sign charges $m$, such that 
    \begin{align}
      (m\c\ldots)\to  (m\c m\c m\c -m\c -m\c\ldots)
    \end{align}
    where $-m$ is the $X$-charge of two new chiral fields $\chi_{R1}$, $\chi_{R2}$.
    This is the minimal requirement for a rank-2 neutrino mass matrix for light neutrinos.
    We note that if the original solution satisfies the Diophantine equations, the second will trivially do it as well.
    \item If $m$ is not present in the solution we can add three sets of opposite sign charges $m$, and identify the $X$-charges of three new chiral fields $\chi_{R1}$, $\chi_{R2}$ and $\chi_{R3}$ as $-m$.
 \end{itemize}

As mentioned previously, in order to limit the total number of solutions that cancel the anomaly induced by the additional $U(1)_X$, the following restrictions are taken into account:
\begin{enumerate}
    \item By construction, all new chiral fermions have to be charged under $U(1)_X$, i.e., solutions with vanishing charges are disregarded.
    \item For the chiral fields, the maximal charge allowed (in absolute value) is 30.
    \item Solutions with vector-like fermions are disregarded. %, i.e., the ones containing two opposite charges.
    We emphasize that the first condition in Eq.~\eqref{eq:vertices} together with the requirement of a rank-2 neutrino mass matrix demand at least 2 opposite pairs of charges.
    However, this does not imply vector-like fermions, since the charges $m$ are associated to SM doublets.
    \item At least two charges have to be equal.
    Their corresponding fields are identified with the RHNs.
    \item A second set of at least two equal charges have to exist.
    Their corresponding fields are identified with the heavy left-handed chiral fermions.
    This is due to the need of a neutrino mass matrix for the light neutrinos of at least rank 2.
    \item A third set of three equal charges is required.
    Their corresponding fields are identified with the SM doublets charged under the new $U(1)_X$.
    \item We restrict ourselves to $N \le 9$ fields, with charges satisfying the two Diophantine conditions in Eq.~\eqref{eq:NN3}, and take the minimal charge (in absolute value) to be positive.
    We note that there are no solutions for $N \leq 5$ with at least two sets of equal charges~\cite{Davoudiasl:2005ks, Nakayama:2011dj}.
    \item The numbers of left-handed chiral fermions $\chi_R$ and $\chi_L$ has to be the same, for their masses to be generated by the SSB of the $S$.
    \item The charge assignment may not allow all chiral fields to acquire masses via the SSB.
    We only consider solutions which have extra massless chiral fields all of them acquiring masses through an extra singlet scalar $S'$ with $X$-charge $s'$.
    This implies that these charges cannot be repeated.
    \item We want RHN masses to be generated by tree-level Dirac seesaw.
    That implies that all vertices between $S$, the RHN and the other chiral fields should be forbidden by the symmetries.
    \item In the case of new set of chiral fermions which get masses through $S'$ the lightest fermion must be a viable DM candidate .
    
    \item In order to have a viable type-I  Dirac  seesaw, new heavy chiral fermions have to satisfy the condition in Eq.~\eqref{eq:condition}.
    For type-II Dirac seesaw, the conditions to be imposed appear in Eqs.~\eqref{eq:type2active} and~\eqref{eq:type2dark} for active or dark symmetries, respectively, as will be seen in the following.
\end{enumerate}
Additionally, we note that for a fixed number of chiral fields, different solutions could share the same qualitative behavior.
For example, the second solution in Table~\ref{tab:sltns} is $(1\c 2\c 2\c -3\c -3\c -3\c 4)$ and has to be expanded with the set of opposite charges $(1\c 1\c -1\c -1)$ to get $(1\c 1\c 1\c -1\c -1\c 2\c 2\c -3\c -3\c -3\c 4)$.
That solution and $(3\c 3\c 3\c -5\c -5\c 7\c 7\c -7\c -7\c -7\c 8)$ are equivalent (and therefore the latter is  omitted in the table) in the sense that both contain three RHNs (with charges $\nu=-3$ and $\nu=3$, respectively), two heavy Dirac fermions (with charges $(-1\c 2)$ and $(-5\c 7)$), and a single massless Majorana fermion (with charge 4 and 8).
Therefore, in this case only one solution (the one with the smallest charge in absolute value) is reported in Table~\ref{tab:sltns}.

\begin{table}
  \centering
  \begin{tiny}
  \begin{tabular}{LLR|LLR|LLLLL}
 \toprule
N &               \ell &                 k &          \text{solution}               &                                   \text{GCD} &  \text{extra}&  l &   \nu &   s &              \text{massless} &    s' \\
\midrule
               6 &          (1, -2) &           (-4, 1) &                  (1, -4, -4, 9, 9, -11) &   3 &               (1, 1, -1, -1) &  -4 &   9 & -5 &           (-11) &    22 \\ \midrule
               7 &          (-1, 1) &       (-1, 0, -1) &                (1, 2, 2, -3, -3, -3, 4) &   1 &               (1, 1, -1, -1) &   2 &  -3 &  1 &             (4) &     8 \\
               7 &           (3, 1) &       (-1, -5, 7) &                (2, 2, -4, 7, -8, -8, 9) &   1 &               (-4, -4, 4, 4) &   2 &  -8 &  6 &          (7, 9) &    16 \\ 
               7 &          (-1, 1) &       (-1, 0, -1) &                (1, 2, 2, -3, -3, -3, 4) &   1 &        (-4, -4, -4, 4, 4, 4) &  -3 &   2 &  1 &          (1, 4) &     5 \\ \midrule
               8 &      (0, -3, -1) &  (-3, -5, -6, -4) &            (3, -4, -6, -6, 7, 7, 8, -9) &   1 &               (8, 8, -8, -8) &   7 &  -6 & -1 &            (-9) &    18 \\
               8 &      (-4, -1, 1) &    (-13, -14, -7) &         (1, 1, -2, -4, -4, 10, 11, -13) &   2 &               (-2, -2, 2, 2) &   1 &  -4 &  3 &            (11) &    22 \\
               8 &        (0, 2, 3) &  (-1, -7, -4, -5) &       (1, -5, -5, -10, 11, 11, 11, -14) &   2 &               (1, 1, -1, -1) &  -5 &  11 & -6 &      (-10, -14) &    24 \\
               8 &       (-1, 0, 1) &      (-11, -8, 8) &       (1, 2, 13, 13, -16, -16, -19, 22) &   2 &           (-19, -19, 19, 19) & -16 &  13 &  3 &            (22) &    44 \\ 
               8 &       (0, 1, -2) &   (-1, -2, -4, 1) &             (2, 2, 2, 2, -5, -5, -5, 7) &   2 &        (-8, -8, -8, 8, 8, 8) &  -5 &   2 &  3 &             (7) &    14 \\
               8 &       (-1, 0, 1) &       (-4, -2, 2) &            (1, -2, -2, 4, 5, -7, -7, 8) &   1 &  (-16, -16, -16, 16, 16, 16) &  -7 &  -2 &  9 &            (16) &    32 \\
               8 &      (-1, 0, -1) &       (-2, 1, -1) &           (2, -5, -5, -5, 7, 8, 8, -10) &   1 &        (-2, -2, -2, 2, 2, 2) &  -5 &   8 & -3 &             (2) &     4 \\
               8 &       (0, -1, 0) &  (-2, -1, -2, -1) &          (1, 4, 5, 5, -8, -10, -10, 13) &   1 &  (-15, -15, -15, 15, 15, 15) & -10 &   5 &  5 &            (15) &    30 \\
               8 &      (5, -7, -1) &   (-5, -8, -6, 0) &        (5, 5, 5, 13, -17, -17, -17, 23) &  60 &  (-29, -29, -29, 29, 29, 29) & -17 &   5 & 12 &        (13, 23) &    36 \\
               8 &      (-1, -2, 2) &     (-10, -1, -8) &      (1, 10, 10, 10, -19, -19, -19, 26) &  17 &  (-28, -28, -28, 28, 28, 28) & -19 &  10 &  9 &         (1, 26) &    27 \\ \midrule
               9 &     (1, 2, 1, 2) &  (-5, -3, -6, -3) &          (1, 1, 2, 2, 4, -5, -7, -7, 9) &   1 &               (-5, -5, 5, 5) &   1 &  -7 &  6 &             (9) &    18 \\
               9 &     (-3, -1, -2) &  (-1, -5, -1, -5) &          (1, 2, 2, 2, 2, -5, -5, -8, 9) &   1 &               (-8, -8, 8, 8) &  -5 &   2 &  3 &          (1, 9) &    10 \\
               9 &      (-4, -1, 1) &  (-3, -9, -7, -1) &         (1, 1, 1, 5, 5, -7, -8, -9, 11) &   2 &           (11, 11, -11, -11) &   5 &   1 & -6 &    (-7, -8, -9) &    16 \\
               9 &     (-3, -1, -2) &  (-1, -3, -2, -3) &        (1, -2, 5, 5, 6, -9, -9, -9, 12) &   2 &               (1, 1, -1, -1) &   5 &  -9 &  4 &            (12) &    24 \\
               9 &    (2, -1, 0, 1) &    (0, -3, 4, -4) &    (2, 6, -7, 8, -11, -12, -12, 13, 13) &   1 &           (-11, -11, 11, 11) & -12 &  13 & -1 &          (2, 8) &    10 \\
               9 &       (2, 1, -2) &   (-1, -3, 0, -4) &      (2, 4, -6, 7, -9, -9, 12, 12, -13) &   2 &               (-6, -6, 6, 6) &  -9 &  12 & -3 &  (2, 4, 7, -13) &     6 \\
               9 &    (-1, 2, 3, 2) &  (-1, -3, -2, -5) &       (1, 1, 2, 5, 8, -9, -11, -11, 14) &   1 &               (-9, -9, 9, 9) &   1 & -11 & \boldsymbol{10} &      (14) &    28 \\
               9 &   (1, 0, -2, -1) &  (-6, -5, -4, -5) &      (1, 1, -4, 5, -6, -6, 10, 14, -15) &   1 &               (-4, -4, 4, 4) &   1 &  -6 &  5 &         (5, 14) &    19 \\
 \boldsymbol{ 9} &     (3, 1, 3, 2) &  (-4, -6, -5, -2) &     (2, -5, 6, 9, 9, -12, -12, -13, 16) &   2 &               (6, 6, -6, -6) &   9 & -12 &  3 &              () &   - \\
               9 &      (-1, 0, -2) &    (-4, -3, 4, 3) &  (1, 8, -12, -12, -12, 14, 14, 16, -17) &   2 &           (16, 16, -16, -16) &  14 & -12 & -2 &        (8, -17) &     9 \\
               9 &    (-1, 1, 0, 2) &    (-4, -5, 4, 0) &  (3, 8, -11, -11, 12, 12, -16, -16, 19) &   4 &               (8, 8, -8, -8) &  12 & -16 &  4 &    (3, 19, -11) &    22 \\
               9 &       (-5, 1, 9) &    (-4, 1, 3, -1) &   (1, -2, 7, 11, 11, -15, -15, -24, 26) &   3 &               (7, 7, -7, -7) &  11 & -15 &  4 &    (1, -24, 26) &     2 \\
               9 &    (2, -1, 3, 1) &  (-6, -5, -3, -4) &          (1, 1, 1, 2, 5, -6, -6, -6, 8) &   1 &  (-11, -11, -11, 11, 11, 11) &  -6 &   1 &  5 &       (2, 5, 8) &    10 \\
               9 &        (2, 1, 3) &    (-2, 0, -5, 8) &       (1, -2, -3, -3, -3, 4, 8, 8, -10) &   1 &        (2, 2, 2, -2, -2, -2) &  -3 &   8 & -5 &       (-2, -10) &    12 \\
               9 &   (-2, -3, 0, 1) &   (-2, -4, -3, 0) &        (1, 4, 5, -6, -6, -6, 9, 9, -10) &   1 &        (-3, -3, -3, 3, 3, 3) &  -6 &   9 & -3 &  (1, 4, 5, -10) &     5 \\
 \boldsymbol{ 9} &    (1, 2, 1, -1) &  (-6, -3, -5, -3) &        (1, 2, -6, -6, -6, 8, 9, 9, -11) &   1 &        (-3, -3, -3, 3, 3, 3) &  -6 &   9 & -3 &              () &    - \\
               9 &  (-3, -6, 1, -4) &   (-3, -6, -5, 4) &      (2, -3, -3, -3, -6, 7, 7, 11, -12) &   2 &        (1, 1, 1, -1, -1, -1) &  -3 &   7 & -4 &       (11, -12) &     1 \\
               9 &      (-2, -1, 0) &  (-1, -3, -5, -3) &       (5, 6, 6, 6, -8, -9, -9, -10, 13) &   2 &  (-12, -12, -12, 12, 12, 12) &  -9 &   6 &  3 &            (12) &    24 \\
               9 &    (2, 0, 1, -1) &   (-6, -3, 2, -5) &       (4, 4, 6, 6, -7, -7, -7, -12, 13) &   4 &        (-8, -8, -8, 8, 8, 8) &  -7 &   6 &  1 &             (4) &     8 \\
               9 &      (-5, -2, 3) &   (-1, -2, 3, -4) &      (3, 5, 5, -8, -8, -8, 12, 12, -13) &   3 &        (-4, -4, -4, 4, 4, 4) &  -8 &  12 & -4 &     (3, -13, 5) &    10 \\
               9 &     (-6, -3, -2) &   (-1, -5, 2, -2) &   (5, 5, -8, 11, 11, -14, -14, -14, 18) &  12 &  (-17, -17, -17, 17, 17, 17) & -14 &  11 &  3 &            (18) &    36 \\
\bottomrule
\end{tabular}
\end{tiny}
\caption{Type-I Dirac neutrino masses: Set of charges satisfying the Diophantine equations together with the conditions enumerated in the text, for $N$ extra chiral fermions, featuring Dirac neutrino masses generated by D-6 operators.
The solutions without massless chiral fermions are highlighted with a bold font $N$, while the solution with unconditional stability through $\mathbb{Z}_{10}$ is highlighted with a bold font for the charge of $S$.
}
\label{tab:sltns}
\end{table}
%%%%%%%%%%%%%%%%%%%%%%%%%%%%%%%%%%%%%%%%%%%%%%%%%%

The solutions of the Diophantine equations satisfying all the previously enumerated conditions are shown in Table~\ref{tab:sltns}. 
The solutions for $N$ extra chiral fermions are parametrized as a function of two sets of integers $\ell$ and $k$ (first three columns).
The fourth column shows the charge assignments, whereas the fifth the general common denominator (GCD) of the original solution.
Technically speaking, the solutions were found using the package \texttt{anomalies}\footnote{\url{https://pypi.org/project/anomalies/}}~\cite{Bernal:2021ezl}.
We note that even if most of the solutions contain massless chiral fermion, there are two solutions without ($N$ highlighted in bold).
Regarding these solutions without massless fermions, a few comments are in order:
\begin{itemize}
    \item The first solution corresponds to $N = 9$ and has the charge assignment (6, 6, 6,  $-6$, $-6$, 2, $−5$, 9, 9, $−12$,$−12$,$ −13$, 16).
    In this case, there are two RHNs with charges $\nu = -12$, two $\chi_L$ with charge $l = 9$, and therefore $S$ has to have a charge $s = -\nu - l = 3$.
    Additionally, three states with charges $m = \nu + 2l = 6$ are required.
    As the original solution contains a single state with charge 6, two extra chiral fields with the same charge are added, together with two $\chi_R$ with charge $r = -m = -6$.
    Finally, there are two extra Dirac fermions $(2\c -5)$ and $(-13\c 16)$ that get mass via the SSB by the scalar $S$, each of them being independent DM candidates protected by a residual $\mathbb{Z}_3$ symmetry.
    Other particles (i.e., the SM doublets, the RHNs and the heavy fermionic mediators of the type-I Dirac Seesaw) are neutral under such symmetry.
    \item The second solution also needs $N = 9$ new chiral fermions and corresponds to the charge assignment $(-3\c -3\c -3\c 3\c 3\c 3\c 1\c 2\c -6\c -6\c -6\c 8\c 9\c 9\c -11)$.
    It contains two RHNs with charge $\nu = 9$, and three $\chi_L$ with $l = -6$.
    As none of the three states with charges $m = \nu +2l = -3$ were present,  three extra chiral fermions $\chi_L$ with charges $r = -m = 3$  have been added.
    Finally, two Dirac fermions $(1\c 2)$ and $(8\c -11)$ acquire mass via the scalar $S$ with charge $s = -\nu - l = -3$.
\end{itemize}

All other solutions presented in Table~\ref{tab:sltns} have a number of massless chiral fermions.
They can be either extra relativistic degrees of freedom, or additional DM candidates if they acquire mass from another mechanism.

Concerning the solutions with multicomponent DM, we explore the cases which feature at least two DM candidates with \emph{unconditional} stability~\cite{Yaguna:2019cvp}. This happens when there are two remnant symmetries such that $\mathbb{Z}_{|s|}\cong \mathbb{Z}_p\otimes\mathbb{Z}_q$ with  $\mathbb{Z}_p\otimes\mathbb{Z}_q$ coprimes, which guaranteed the stability of each lightest state under $\mathbb{Z}_p$ and $\mathbb{Z}_q$ respectively, without imposing any kinematical restriction. For the two DM candidates associated to the set of chiral fields $\psi_i$ and $\chi_j$, we consider the first two possibilities for $|s|$~\cite{Yaguna:2019cvp}
\begin{itemize}
    \item $\mathbb{Z}_{6}\cong \mathbb{Z}_2\otimes\mathbb{Z}_3$: solutions with at least a set of chiral fields with $\psi_i \sim\left[\omega_{6}^{2} \vee \omega_{6}^{4}\right]$ under $\mathbb{Z}_{6}$, and at least a set of chiral fields with $\chi_i \sim \omega_{6}^3$ under $\mathbb{Z}_{6}$,
    \item $\mathbb{Z}_{10}\cong \mathbb{Z}_2\otimes\mathbb{Z}_5$: solutions with at least a set of chiral fields with $\psi_i \sim\left[\omega_{10}^{2}  \vee \omega_{10}^{6} \vee \omega_{10}^{8} \right]$ under $\mathbb{Z}_{10}$ and at least a set of chiral fields with $\chi_i \sim \omega_{10}^5$ under $\mathbb{Z}_{10}$,
\end{itemize}
where $\omega_{|s|}=\operatorname{e}^{i\, 2\pi/|s|}$. The solutions with unconditional stability are highlighted with a bold font in the column $s$ of Table~\ref{tab:sltns}. In this case, we have only a $\mathbb{Z}_{10}$ solution in which we have a first set of singlet chiral fermions $\psi_i$ with charges $(\omega_{10}^2,\, \omega_{10}^8)$ and a second  singlet chiral fermion $\chi$ with charge $\omega_{10}^{5}$. This give to arise a Dirac fermion DM candidate $\begin{pmatrix}
       \psi_1 & \left(\psi_2\right)^\dagger
\end{pmatrix}^{\operatorname{T}}$ protected by $\mathbb{Z}_{5}$ and a Majorana fermion DM   $\begin{pmatrix}
       \chi & \left(\chi\right)^\dagger
\end{pmatrix}^{\operatorname{T}}$ protected by $\mathbb{Z}_{2}$.

Finally, the type-I Dirac seesaw realizations of the effective operator of D-6 in Eq.~\eqref{eq:nmo56} have to have a sufficiently rich $h_{\nu}^{\alpha i}$ structure to explain the full neutrino oscillation data.
That can be guaranteed by having a rank 2 or 3 Dirac neutrino mass matrix, via the inclusion of a proper set of chiral fermions for each solution.
For example, consider the first solution $(1\c -4\c -4\c 9\c 9\c -11)$, which is promoted to $(1\c 1\c 1\c -1\c -1\c -4\c -4\c 9\c 9\c -11)$ by adding two pairs of extra states $(1, -1)$. We assign $m=1$, $r=-1$, $l=-4$ and $\nu=9$, such that $s=-5$ gives masses to two Dirac neutrinos.
For avoiding the chiral fermion with charge $-11$ to be massless, an extra scalar $S'$ with $X$-charge $22$ can be introduced to give a Majorana mass.

%%%%%%%%%%%%%%%%%%%%%%%%%%%%%%%%%%%%%%%%%%%%%%%%%%
\subsection{Type-II Dirac Seesaw}
\begin{figure}
  \def\scl{0.65}
  \def\sepf{4.4cm}
  \centering
  \includegraphics[scale=\scl]{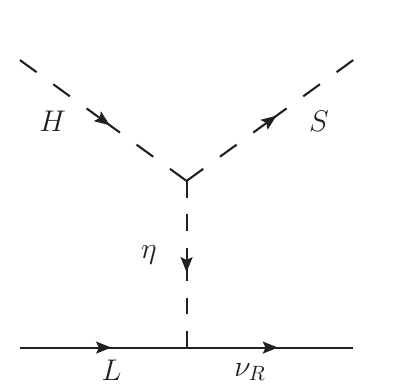}\qquad
  \includegraphics[scale=\scl]{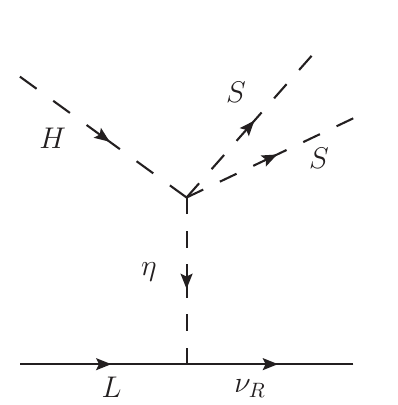}\\
  \quad \texttt{\hspace{-0.8cm} \scriptsize D-5 \hspace{\sepf} D-6} \\
  \caption{Diagrams leading to tree-level type-II Dirac neutrino masses, via the dimension 5 (left panel) or dimension 6 (right panel) operators described in Eq.~\eqref{eq:nmo56}.
 }
  \label{fig:topoII}
\end{figure}

Contrary to type-I, type-II Dirac seesaw with only chiral fermions can be realized via D-5 operators.
Figure~\ref{fig:topoII} presents the diagrams realizing the D-5 (left panel) and D-6 (right panel) effective Lagrangian in Eq.~\eqref{eq:nmo56} for type-II seesaw, if one only allows SSB masses for the singlet chiral fermions.

%%%%%%%%%%%%%%%%%%%%%%
%%%%%%%%%%%%%%%%%%%%%%
For the case D-5, the flux of the $X$-charges in each vertex for an {\it active} symmetry satisfy
\begin{align} \label{eq:vertices2}
    L-m=&\eta+s\,,& -L+\eta =&\nu \,,
\end{align}
which implies that
\begin{equation}\label{eq:type2active}
    \nu+m+s=0  \,.
\end{equation}
For D-6, the type-II seesaw simply implies a change of charge of $s$ to $s/2$, as in the upper vertex there are two ongoing $S$ instead of a single one.

\subsubsection{Active Symmetry}
The solutions of the Diophantine equations satisfying all the previously enumerated conditions for a type-II Dirac neutrino masses and an active symmetry, are shown in Table~\ref{tab:sltns2} for the D-5 operator. 
%%%%%%%%%%%%%%%%%%%%%%%%%%%%%%%%%%%%%%%%%%%%%%%%%%
\begin{table}
  \centering
  \begin{scriptsize}
  \begin{tabular}{LLR|LR|LLLLL}
 \toprule
N &               \ell &                 k &                                                    \text{solution} & \text{GCD} &  m &   \nu &   s &              \text{massless} &    s' \\
\midrule
              6 &         (-1, -2) &           (-1, 2) &                (1, 1, 1, -4, -4, 5)\text{~\cite{Gu:2019yvw}} &   1 &  1 &  -4 &  3 &             (5) &    10 \\ \midrule
              7 &          (-1, 1) &       (-1, 0, -1) &            (1, 2, 2, -3, -3, -3, 4) &   1 & -3 &   2 &  1 &          (1, 4) &     5 \\ \midrule
              8 &      (1, -3, -2) &  (-4, -9, -5, -3) &         (1, 3, 3, 3, -5, -7, -7, 9) &   1 &  3 &  -7 &  4 &             (9) &    18 \\
              8 &     (-8, -1, -4) &   (-2, -6, 4, -5) &  (7, -8, -18, -18, 20, 20, 20, -23) &   2 & 20 & -18 & -2 &    (-23, -8, 7) &    16 \\ \midrule
              9 &    (3, 0, -1, 1) &   (-6, -5, 5, -1) &     (2, 2, 2, -3, -3, 4, -5, -5, 6) &   1 &  2 &  -5 &  3 &             (4) &     8 \\
              9 &  (1, -1, -2, -1) &  (-6, -5, -3, -5) &      (1, 1, 2, 2, 2, -3, -6, -8, 9) &   1 &  2 &   1 & -3 &        (-8, -3) &    11 \\
              9 &   (-2, -3, 0, 1) &   (-2, -4, -3, 0) &    (1, 4, 5, -6, -6, -6, 9, 9, -10) &   1 & -6 &   9 & -3 &  (-10, 1, 4, 5) &     5 \\
 \boldsymbol{9} &    (1, 2, 1, -1) &  (-6, -3, -5, -3) &    (1, 2, -6, -6, -6, 8, 9, 9, -11) &   1 & -6 &   9 & -3 &              () &  - \\
              9 &      (-5, -2, 3) &   (-1, -2, 3, -4) &  (3, 5, 5, -8, -8, -8, 12, 12, -13) &   3 & -8 &  12 & -4 &     (-13, 3, 5) &    10 \\
\bottomrule
\end{tabular}
\end{scriptsize}
\caption{Type-II Dirac neutrino masses for an {\it active} symmetry: Set of charges satisfying the Diophantine equations together with the conditions enumerated in the text, for $N$ extra singlet chiral fermions, featuring Dirac neutrino masses generated by D-5 operators.
The solutions without massless chiral fermions are highlighted with a bold font.}
\label{tab:sltns2}
\end{table}
%%%%%%%%%%%%%%%%%%%%%%%%%%%%%%%%%%%%%
We note that even if most of the solutions contain massless chiral fermions, there is a single solution without (highlighted in bold).
It corresponds to a case with $N = 6$ new chiral fermions, with the charge assignment ($1\c 2\c −6\c −6\c −6\c 8\c 9\c 9\c −11$).
It contains two RHNs with charge $\nu = 9$, and three states with charge $m = -6$.
The other four chiral fermions form two Dirac states $(1\c 2)$ and $(9\c -11)$ that obtain mass via the scalar $S$ with charge $s = -\nu - m = -3$, and that could be viable DM candidates.

\subsubsection{Dark symmetry}
Contrary to the type-I seesaw, the type-II seesaw can accommodate a {\it dark} symmetry~\cite{Ma:2021szi}.%
\footnote{We note that this case reduces to the one with an active symmetry in the limit $L = 0 = m$.}
For the case D-5, the flux of charges in each vertex is
\begin{align} \label{eq:vertices3}
    0=&\eta+s\,,& \eta =&\nu \,.
\end{align}
In general, for realizing the Type-II Dirac Seesaw, one requires
\begin{align}\label{eq:type2dark}
    \nu+m+s/\delta=0\,,
\end{align}
with $m$ for a gauge $U(1)_D$ symmetry, $\delta=1$ or $2$ for D-5 or D-6 realizations, as in the upper vertex there are two ongoing $S$ instead of a single one.

The solutions of the Diophantine equations satisfying all the previously enumerated conditions for a type-II Dirac seesaw neutrino mass mechanism and an dark symmetry, are shown in Table~\ref{tab:sltns3}, for the D-5 operator.
A total of 19 solutions where found, 3 of them (highlighted in bold) without massless fermions.%
\footnote{The first two solutions were recently presented in Ref.~\cite{Ma:2021szi}.}
\begin{table}
  \centering
\begin{scriptsize}
\begin{tabular}{LLR|LR|RRRR}
 \toprule
N &                \ell &                 k &                           \text{solution} & \text{GCD} &  \nu &  s &\text{massless} & s'  \\
\midrule
              6 &          (-1, -2) &           (-1, 2) &               (1, 1, 1, -4, -4, 5)\text{\cite{Ma:2021szi}} &   1 &   1 &  -1 &           (-4) &    -8 \\
 \boldsymbol{6} &           (-1, 1) &           (-2, 0) &              (1, -2, -3, 5, 5, -6)\text{\cite{Ma:2021szi}} &   1 &   5 &  -5 &             () &  - \\
              6 &            (0, 2) &       (-1, -2, 1) &            (1, 1, 8, -11, -16, 17) &   1 &   1 &  -1 &       (-11, 8) &     3 \\ \midrule
              7 &           (-1, 1) &      (-1, -2, -1) &           (1, 3, -4, 5, -6, -6, 7) &   1 &  -6 &   6 &        (-4, 7) &     3 \\
              7 &       (-1, 0, -2) &       (-1, 0, -1) &          (1, 7, 8, -9, -9, -9, 11) &   1 &  -9 &   9 &        (7, 11) &    18 \\ \midrule
              8 &       (-1, 2, -2) &        (-7, 4, 0) &        (1, 2, 2, 2, -3, -5, -6, 7) &   1 &   2 &  -2 &           (-6) &    12 \\
              8 &        (3, 2, -2) &     (-7, -10, -4) &       (2, -3, -4, 5, -6, 7, 7, -8) &   1 &   7 &  -7 &       (-8, -6) &    14 \\
              8 &        (1, -2, 2) &    (-1, 0, -3, 5) &    (4, -5, -5, 7, 8, -10, -10, 11) &   1 & -10 &  10 &  (4, 7, 8, 11) &    15 \\
              8 &       (-2, -1, 2) &       (-2, -5, 1) &     (1, 2, 2, -7, -7, 10, 10, -11) &   2 &  10 & -10 &        (-7, 2) &     5 \\ \midrule
              9 &  (-2, -3, -1, -2) &   (-2, -6, -5, 3) &(3, -4, -6, -6, -8, 10, 14, 16, -19)&   1 &  -6 &\boldsymbol{6} &(-19, 16) & 3 \\
              9 &      (1, 0, 4, 3) &   (-2, 0, -2, -4) & (1, -3, 6, 6, 6, -7, -10, -15, 16) &   1 &   6 &\boldsymbol{-6} &(-15) &    30 \\              
              9 &  (-6, -3, -6, -2) &  (-6, -4, -1, -5) &     (1, 1, 2, 2, 3, -5, -6, -6, 8) &   1 &   2 &  -2 &           (-6) &   -12 \\
 \boldsymbol{9} &   (-2, -3, 1, -1) &  (-3, -1, -2, -1) &    (1, -2, 3, 4, 6, -7, -7, -7, 9) &   1 &  -7 &   7 &             () &  - \\
              9 &       (-3, -1, 5) &   (-9, 3, -4, -1) &    (1, 2, -3, 4, -5, -6, 8, 8, -9) &   1 &   8 &  -8 &        (-6, 2) &     4 \\
              9 &       (-2, 1, -2) &   (-1, -2, -1, 3) &    (2, 2, 4, -5, -5, -5, 8, 8, -9) &   1 &  -5 &   5 &         (2, 8) &    10 \\
              9 &        (4, 5, -1) &  (-2, -1, -3, -2) &(2, -5, 8, 10, 10, -12, -16, -18, 21)&  1 &  10 &\boldsymbol{-10}&(-16, 21) & 5 \\
 \boldsymbol{9} &        (-2, 0, 2) &    (-1, 1, 0, -1) &  (1, 1, -4, -5, 9, 9, 9, -10, -10) &   1 &   9 &  -9 &             () &  - \\
              9 &     (2, -2, 1, 3) &  (-5, -4, -3, -1) &  (1, -2, -2, -2, 5, -7, 8, 9, -10) &   1 &  -2 &   2 &            (9) &    18 \\
              9 &    (-2, -3, 0, 1) &   (-2, -4, -3, 0) &   (1, 4, 5, -6, -6, -6, 9, 9, -10) &   1 &  -6 &   6 &            (9) &    18 \\
              9 &       (-4, -5, 3) &   (-2, 0, -1, -2) &   (3, 3, -4, 5, 5, -6, -8, -8, 10) &   1 &  -8 &   8 &       (-6, 10) &     4 \\
              9 &      (1, 0, 2, 1) &   (-5, 3, -3, -6) &  (1, -3, -3, -4, -5, 8, 9, 9, -12) &   1 &  -3 &   3 &            (9) &    18 \\
              9 &     (-2, 0, 1, 3) &    (-1, 0, -1, 3) &  (1, 1, 2, 2, -6, 9, -10, -12, 13) &   1 &   1 &  -1 &        (-6, 2) &     4 \\
              9 &         (3, 0, 5) &  (-2, -5, -4, -2) &(1, 5, 9, -10, -10, -10, 16, 25, -26)&  3 & -10 &\boldsymbol{10} &(25) &    50 \\
\bottomrule
\end{tabular}
\end{scriptsize}
\caption{Type-II Dirac neutrino masses for a {\it dark} symmetry: Set of charges satisfying the Diophantine equations together with the conditions enumerated in the text, for $N$ extra singlet chiral fermions, featuring Dirac neutrino masses generated by D-5 operators.
None of the present solutions contain massless chiral fermions.
}
\label{tab:sltns3}
\end{table}
Regarding these solutions without massless fermions, a few comments are in order:
\begin{itemize}
    \item The first solution corresponds to $N = 6$ and has the charge assignment $(1, -2, -3, 5, 5, -6)$.
    It contains two RHNs with charge $\nu = 5$.
    There are also two extra Dirac fermions $(1\c -6)$ and $(-2\c -3)$ that get mass via the SSB by the scalar $S$ with charge $s = -\nu = -5$, each of them being independent DM candidates protected by a residual $\mathbb{Z}_5$ symmetry.
    \item The other two solutions contain $N = 9$ new chiral fermions, the first with the assignment $(1, -2, 3, 4, 6, -7, -7, -7, 9)$.
    It contains three RHNs with charge $\nu = -7$, and three Dirac fermions $(1, 6)$, $(-2, 9)$ and $(3, 4)$ gaining mass via a scalar of charge $s = 7$.
    \item Finally, the last solution has charges $(1, 1, -4, -5, 9, 9, 9, -10, -10)$.
    It contains three RHNs with charge $\nu = 9$, and three Dirac fermions $(1, -10)$, $(1, -10)$ and $(-4, -5)$ that obtain mass via the scalar $S$ with charge $s = -9$.
    Additionally, the scalar $\eta$ has charge $\eta = \nu = 9$.
    The dark $U(1)_D$ symmetry is broken by $S$ down to a $\mathbb{Z}_9$ symmetry.
  \end{itemize}
We also have four solutions with unconditional stability. Concerning the second with $\mathbb{Z}_6 \cong \mathbb{Z}_2 \otimes \mathbb{Z}_3 $, 
some phenomenological considerations are in order: the particle content of the model, along with the charges of the remnant symmetries, are presented in Table~\ref{tab:pickedsltn}. The interaction Lagrangian includes
\begin{align}
  {\cal L}_{\text{int}}\supset y^{\chi}_{ab}\chi \chi S^{*} + y^{\psi}\psi_1 \psi_2 S + y^{\xi} \xi_1 \xi_2 S^* 
+h_{\alpha i} \left( \nu_{R\alpha} \right)^{\dagger} L_{i}\cdot  \eta+\text{H.c.}-V(\eta,H,S)\,,
\end{align}
where
\begin{align}
  V\left(\eta,H,S  \right)\supset&\mu^2 H^{\dagger}H+\lambda \left( H^{\dagger}H \right)^2+\mu^2_{\eta} \eta^{\dagger}\eta+
  \lambda_{\eta} \left( \eta^{\dagger}\eta \right)^{\dagger}+\mu_S^{2} S^{*} S +\lambda_S \left(  S^{*} S \right)^2 \nonumber\\
+&\lambda_{SH} S^{*}S H^{\dagger}H+\lambda_{\eta H} \eta^{\dagger}\eta H^{\dagger} H +\lambda_{S\eta} S^{*}S \eta^{\dagger}\eta 
   +\left[\kappa H^{\dagger} \eta S +\text{H.c.} \right]\,,
\end{align}
and
\begin{align}
 \langle H \rangle =&\frac{v}{\sqrt{2}}\,,&  \langle \eta \rangle =&\frac{v_{\eta}}{\sqrt{2}}\,,& \langle S\rangle=&\frac{v_{S}}{\sqrt{2}}\,.
\end{align}

\begin{table}
  \centering
  \begin{tabular}{l|c|c|c|c|c|c}\hline
    Field&$SU(2)_L$ & $U(1)_Y$&$U(1)_D$&$\mathbb{Z}_6$&$\mathbb{Z}_2$&$\mathbb{Z}_3$\\\hline
     $L_i$      & $\mathbf{2}$ & $-1$   & $0$    & $1$  & $+$  & $1$     \\ %$e^{2\pi 9/9}=e^{2\pi}$
     $\nu_{R \alpha}$  & $\mathbf{1}$ & $0$   & $6$   & $1$  & $+$  & $1$     \\
     $\psi_1$       & $\mathbf{1}$ & $0$  & $-10 $ & $\omega_{6}^2$& $+$& $\omega_3^2$   \\ %$e^{2\pi (-4-5+5)/9}=e^{-2\pi}e^{2\pi 5/9}=e^{2\pi (3)/9}$
     $\psi_2$      & $\mathbf{1}$ & $0$   & $16$   & $\omega_{6}^4$& $+$& $\omega_3$   \\ %$e^{2\pi (-5-4+4)/9}=e^{-2\pi}e^{2\pi 4/9}$
     $\chi$      & $\mathbf{1}$ & $0$    & $-3$   & $\omega_{6}^{3}$& $-$& $1$   \\
     $\chi'$      & $\mathbf{1}$ & $0$   & $-15$  & $\omega_{6}^{3}$& $-$& $1$   \\
     $\xi_1$     & $\mathbf{1}$ & $0$    & $1$    & $\omega_{6}$ & $-$ & $\omega_3$   \\ %$e^{2\pi (-18+8)/9}=e^{-4\pi}e^{2\pi 8/9}$
     $\xi_2$     & $\mathbf{1}$ & $0$    & $-7$   & $\omega_{6}^5$& $-$& $\omega_3^{2}$   \\ %$e^{2\pi (-18+8)/9}=e^{-4\pi}e^{2\pi 8/9}$
     $H$       & $\mathbf{2}$ & $1$ &  $0 $  & $1$  & $+$  & $1$    \\        
     $\eta$       & $\mathbf{2}$ & $1$ &  $6 $  & $1$  & $+$  & $1$    \\        
     $S$       & $\mathbf{1}$ & $0$   &  $-6$  & $1$  & $+$  & $1$    \\\hline %$e^{2\pi (-9)/9}=e^{-2\pi}=1$
  \end{tabular}
  \caption{Charges for last solution. $i=1,2,3$, $\alpha=1,2,3$. Note that $\left( \omega_n^d \right)^{*}=\omega_n^{-d}=\omega_n^{n-d}$.}
  \label{tab:pickedsltn}
\end{table}

After the spontaneous symmetry breaking, the neutrino mass matrix reads
\begin{align}
  {\cal M}_{\alpha i} = \frac{1}{\sqrt{2}}\, h_{\alpha i}\, v_{\eta}\,,
\end{align}
and therefore, $v_{\eta}$ must be small to allow for sizeable Yukawa couplings. This condition can be easily satisfied if $v_{\eta}\ll v$.
In fact
\begin{align}
  v_{\eta} \approx \kappa \frac{v v_{S}}{2 M_{\eta}^{2}}
\end{align}
is expected to be much more smaller than $v$ for large $M_{\eta}$.

Contrary to the scotogenic model, the dark sector is completely independent of the heavy particles associated to the type-II Dirac seesaw.
Moreover, because the $\mathbb{Z}_6$ symmetry has two subgroups, the subsequent two dark sectors are completely independent between them, and therefore, we have at least two independent DM candidates. The first one, protected by the subgroup $\mathbb{Z}_{3}$, is the Dirac fermion $\psi=
\begin{pmatrix}
  \psi_1 & \left( \psi_{2} \right)^{\dagger}
\end{pmatrix}^{\operatorname{T}}
$, while the second one, protected by the subgroup $\mathbb{Z}_2$, is the Majorana fermion $\Xi=
\begin{pmatrix}
  \chi & \left( \chi \right)^{\dagger}
\end{pmatrix}^{\operatorname{T}}
$. For simplicity, we assume that other potential DM candidates are heavier than $\psi$ and $\Xi$ and have very small densities in the early universe, primarily by DM conversion into $\psi$ and $\Xi$ mediated by $Z'$ and $S$~\cite{Wong:2020obo}. Besides the DM conversion processes, the proper relic density for the Majorana DM candidate through the annihilation  $\Xi \Xi\to SS$ ($m_{S}<|m_{\Xi}|$), and compatible with direct detection constraints from PandaX-4T~\cite{Meng:2021mui}, was analyzed for the first model in Table~\ref{tab:sltns3} in Ref.~\cite{Ma:2021szi}. The authors did the same analysis for the second model in Table~\ref{tab:sltns3} with one of Dirac DM candidates through the annihilation  $\psi\overline{\psi}\to Z_{D}Z_{D}$ ($m_{D}<|m_{\psi}|$).

We can also introduce an extra singlet scalar in the dark sector with a $D$-charge different from $\pm 6$, $\phi$, to allow the Yukawa coupling between one of the heavy chiral fields with the right-handed neutrinos~\cite{Guo:2020qin}, as for example  $y_{\phi\alpha} \psi_1 \nu_{R\alpha} \phi$ for a $\phi$ with $D$-charge $4$. This coupling leads to a Dirac neutrino portal scenario which can explain fermion DM in the context of the type-II Dirac seesaw~\cite{Biswas:2021kio}.

%%%%%%%%%%%%%%%%%%%%%%%%%%%%%%%%%%%%%%%%%%%%%%%%%%%%%%%%%
%%%%%%%%%%%%%%%%%%%%%%%%%%%%%%%%%%%%%%%%%%%%%%%%%%%%%%%%%
\section{Conclusions} \label{sec:con}

Studies on tree-level Dirac neutrino masses have typically focused on finding specific anomaly-free solutions for a given kind of symmetry, either for an active or dark symmetry. Alternatively, in the present work a complete set of relevant anomaly-free solutions to the general problem of the generation of Dirac neutrino masses at tree level with chiral singlet fermions has been presented. In particular, we restricted the analysis to solutions satisfying a set of general conditions enumerated in the text.
An intensive scan was performed, looking for solutions generating neutrino masses by the type-I and type-II tree-level Dirac seesaw mechanism, via operators with dimension 5 and 6, that correspond to active or dark symmetries.
Each of the presented solutions leads to a unique model with specific phenomenological implications.

It is interestingly to note that type-I Dirac seesaw can only take place for active symmetries, if one demands all extra fermions to be charged under the new symmetry.
Additionally, type-I Dirac seesaw with dimension-5 operators automatically implies a vector-like pair of singlet fermions, and is therefore not possible with only chiral fermions.
However, for dimension-6 operators we found a set of 36 solutions of the Diophantine equations (i.e. anomaly-free solutions) satisfying general conditions enumerated in the text (see Table~\ref{tab:sltns}).
Among them, only 2 solutions with all extra fermions getting mass via the spontaneous symmetry breaking of the new Higgs field.
The massless fermions of the other solutions can either contribute to the relativistic degrees of freedom $\Delta N_\text{eff}$ in the early universe~\cite{Calle:2019mxn}, or acquire masses after the introduction of an extra singlet scalar, becoming independent DM candidates~\cite{Bernal:2018aon}.

Contrary to type-I, type-II Dirac seesaw with only chiral fermions can be realized via dimension-5 operators.
For the case of an active symmetry and dimension-5 operators, 9 solutions were found, only one of them without massless chiral fermions (see Table~\ref{tab:sltns2}).
Alternatively, for the case of a dark symmetry, 19 solutions were found, 3 of them without massless chiral fermions (see Table~\ref{tab:sltns3}).

Most of the solutions found could feature multicomponent DM.
Special attention was brought to those with at least two DM candidates with unconditional stability, which guarantees the viability of DM without imposing any kinematical restriction.

%%%%%%%%%%%%%%%%%%%%%%%%%%%%%%%%%%%%%%%%%%%%%%%%%%%%%%%%%
%%%%%%%%%%%%%%%%%%%%%%%%%%%%%%%%%%%%%%%%%%%%%%%%%%%%%%%%%
\section*{Acknowledgments}
NB received funding from the Spanish FEDER / MCIU-AEI under grant FPA2017-84543-P, and the Patrimonio Autónomo - Fondo Nacional de Financiamiento para la Ciencia, la Tecnología y la Innovación Francisco José de Caldas (MinCiencias - Colombia) grant 80740-465-2020.
The  work  of  DR  is  supported  by  Sostenibilidad  UdeA,  the  UdeA/CODI  Grants 2017-16286 and 2020-33177.
This project has received funding/support from the European Union's Horizon 2020 research and innovation programme under the Marie Skłodowska-Curie grant agreement No 860881-HIDDeN.

\bibliographystyle{apsrev4-1long}
\bibliography{biblio}

\end{document}